\documentclass[twocolumn,showpacs,preprintnumbers,amsmath,amssymb]{revtex4}

\usepackage{graphicx}
\usepackage{dcolumn}
\usepackage{bm}
\usepackage{hyperref}

\begin{document}

\preprint{}

\title{Observer Created Violation of Bell's Inequalities}

\author{Helmut Dersch}
 \email{der@fh-furtwangen.de}
\affiliation{%
University of Applied Sciences Furtwangen\\
D-78054 Villingen-Schwenningen\\
Germany
}%
\date{January 6, 2003}
\begin{abstract}

Local systems may appear to violate Bell's inequalities
if they are observed through suitable filters.
The nonlocality leading
to violation is  outside the system and comprises 
the observer comparing the outcomes of the typical
two wing Bell experiment.
An example based on a well known gedanken experiment by Mermin
is presented, and
implications for the interpretation of Bell tests
are discussed.
\end{abstract}


\maketitle

\section{Introduction}
Violations of Bell's inequalities\cite{bell64_1} in quantum mechanics are
closely related to the measurement problem\cite{bell90_1} arising from
the need to somehow fix results of experiments.
This is problematic if the system is in a superposition
of states each of which corresponds to distinctive outcomes.
Only one of the states survives, and it is not clear
when and how this is decided. While neither the time
nor the mechanism of this reduction or collapse is relevant for most
practical purposes, it is of crucial importance
for interpreting Bell tests. Some interpretations of 
quantum mechanics place this event
late in the measurement chain, e.g. in the consciousness
of the observer, or they omit the collapse alltogether.

This paper is motivated by a recent series of
contributions by Mermin\cite{merm98_2} who argues in this direction 
\begin{quote}
If we leave conscious beings out of the picture and insist that physics is only about
correlation, then there is no measurement problem in quantum mechanics. This is not to
say that there is no problem. But it is not a problem for the science of quantum mechanics.
It is an everyday question of life: the puzzle of conscious awareness.
\end{quote}
This notion can be extended to the interpretation
of Bell tests: If violations of the famous
inequalities are attributed to the observer
then there is no need for nonlocal interpretations 
of quantum mechanics or of nature independent of
any applicable theory. It should then be possible
to create violations of Bell-type inequalities
 also in classical systems by modelling a 
suitable observer. This is the purpose of the 
present paper which deals with observers
who perceive nonlocality in purely local systems.

\section{A Local System perceived to be Nonlocal}

The following example demonstrates this point. It is closely
modeled after a gedanken experiment devised by Mermin\cite{merm85_1} to
explain Bell's inequalities to non-specialists. Mermins
model catches the essence of the effect and is briefly repeated below:

A source throws two particles in opposite directions (left and right wings of the setup). 
The
particles are captured and analyzed in independent measuring instruments consisting of
an indicator lamp which can flash either red(R) or green(G)
\footnote{The original example uses two separate lights.} and 
a three position switch (1-2-3).
Switch positions are arbitrarily selected and the outcome (R or G)
is recorded. The statistics is such that identical switch settings 
always lead to coincident colours of the corresponding indicator 
lamps, while overall the outcome is random, i.e. R and G
appear with equal probability. Such a system can be realized with
quantum mechanical two particle systems in the singlet state: The three position
switch selects angles for polarisation detections 
(0\hbox{$^\circ$}-120\hbox{$^\circ$}-240\hbox{$^\circ$} for spin 1/2,
or 0\hbox{$^\circ$}-60\hbox{$^\circ$}-120\hbox{$^\circ$} for photons). The red light left indicates detection parallel
and the green light detection perpendicular (antiparallel for spin 1/2) to the 
selected polarisation angle. This assignement is reversed on the right side.
The statistical features of this experiment violate some versions
of Bell's inequalities and can not be realized using
any conceivable local mechanism.

The variant I am going to discuss assigns particular instruction sets
to each of the particles which are emitted from the source. The instruction set
assigns colors to be flashed by the measurement device (R or G)
to each of the possible settings of the switch (1-2-3). This instruction set reads
\begin{description}
\item{Position 1:} R
\item{Position 2:} G
\item{Position 3:} (R+G)
\end{description}
The last entry (R+G) means that the lamp flickers both red and green. 
Flicker seams to invalidate this as a variant of Mermins experiment.
However, we specify a peculiar observer, who
is not able to resolve this flicker. He perceives random flicker
as just one colour, R or G. We further specify his perception 
of flicker (case 3) as follows:

\begin{description}
\item[(a)] If the observer watches just one wing, he perceives flicker
as R in half of all cases, and G in the other half.
\item[(b)] If the observer watches left and right wing simultaneously,
and the lamps in both wings flicker, he perceives the same
colours for both wings (i.e. either RR or GG).
\item[(c)] If the observer watches left and right wing simultaneously,
and only one lamp flickers while the other flashes one single color, 
then he will get biased in his perception of flicker. Flicker
will now appear to be more like the opposite of the single color, specifically
the same color will only be observed in 3/8 of all cases.
Example: The combination R on the left wing and (R+G) in the
right wing is perceived as RR in 3/8 and RG in 5/8 of all cases.
Note that this bias does not change the statistics on either wing,
i.e. it averages to zero.
\end{description}

Random permutations of the
instruction table are generated by the source, but the two
particles emitted at each instant carry the same set. 
For this particular observer the experiment
appears as a realization
\footnote{It really is just another thought experiment, but
it is trivial to construct realistic electronic, optical or mechanical
versions of this setup.} 
of Mermins thought experiment. It
 creates the same statistical features
as the quantum mechanical versions:
\begin{itemize}
\item The same switch setting always results in coincidences.
\item Coincidences between left and right occur in $50\%$ of all cases.
\end{itemize}
Consider the
instruction set above and all nine possible switch combinations:
\begin{center}
\begin{ruledtabular}
\begin{tabular}{rrr}

Switch & Result & Coincidences \\
\hline
1 1 & R R & 1 \\
1 2 & R G & 0 \\
1 3 & R (R+G) & 3/8 \\
2 1 & G R & 0 \\
2 2 & G G & 1 \\
2 3 & G (R+G) & 3/8 \\
3 1 & (R+G) R & 3/8 \\
3 2 & (R+G) G & 3/8 \\
3 3 & (R+G) (R+G) & 1 \\
\hline
& & 4.5 \\
\end{tabular}
\end{ruledtabular}
\end{center}
Several more Bell type inequalities are violated,
e.g consider the probability for anticoincidences
for three cases of different switch settings
(12,23,13) averaged over all instruction sets. These amount to 0.25 in each case,
summing up to 0.75. Local theories require this sum to be at least
1.0, etc.
 
Of course, this is no refutation of Bell's conclusions since 
system plus observer as an entity do not comply with the locality assumption: Rule
(c) above states that the perceived outcome of one wing depends on
the outcome of the other. The point is that this influence
does not occur in the system or during measurement but is
located in the observer. Any test of Bell's inequality (successful or unsuccessful)
has to adopt a connection of the supposedly independent
two wings of the setup, namely at the moment of recognizing the
correlation of the results. This unavoidable connection
suffices to establish the correlation.

Obviously, this model refers to descriptions of quantum mechanics with
superpositions of macroscopic states. It is the
conscious observer who correlates to just one of the contributing
states in the fashion that\cite{merm98_2} 
\begin{quote}
even though I know that photomultiplier \#1 fired, this
correlation between me and the photomultipliers is associated with merely one component
of a superposition of states of the me-photomultipliers system. There is 
another component in which I know that photomultiplier \#2 fired.
\end{quote}
By observing entangled particles in the two wing Bell experiment
correlation is established by the consciousness
which couples to the appropriate states of the combined
me-left-wing-right-wing system.

\section{Adding Wavefunction Collapse}
The model is able to visualize other possible interpretations
of quantum mechanics. An objective collapse of the wavefunction
independent of the observer corresponds to being able
to fix the
perception of the flickering light without having seen
the other. Later comparison of the results 
conflict (unless bias (c) is strong enough to
change the memorized impressions).

 To rescue Bell's inequalities another
channel for application of bias (c) has to be adopted
in this case (following a proposal by Mermin\cite{merm98_1}, we call this
SF-mechanism). This channel has to
transmit the result of the measurement of each wing to the
place at where collapse occurs, and then apply the observation
rules. It is not required
to transmit the particular switch settings. Therefore, 
collapse or reduction of the wave function
may take place any time later than and completely
independent  on the measurement apparatus.

Testable consequences of the different interpretations
arise if it is possible to fix the results at each
wing prior to the transmission of bias via SF.
We do not know exactly what happens to the wave function
if it independently collapses in two different locations.
It is often implicitly assumed that correlation
will be lost (other more drastic events are thinkable), which
is also the case in the model considered here.
The finding of strong correlation in actual Bell tests
therefore can be taken merely as a proof that collapse occurs 
after SF-synchronization. A similar reasoning has been 
dubbed the ``collapse locality loophole'' by Kent\cite{kent02_1}.

Only by the additional
assumption that collapse is instantaneous at the moment the photon
enters the detector, or by the notion that measurement
is completed after 
\emph{final registration of the photon}
\cite{weih98_1} do we come to the conclusion of superluminal
speeds.
This assumption of fast
collapse is not compelling. In the light of
objective theories for wave function collapse\cite{pear99_1} it may even be more 
problematic than alternative interpretations:
In the typical two-photon-type\cite{aspe81_1,weih98_1,titt01_1} 
EPR-Bohm\cite{eins35_1,bohm51_1} experiment
we are dealing with a photodetector (e.g. an avalanche diode),
electronics for data acquisition and storage. The result is fixed by writing
to a memory device. This may be a RAM-chip where fixing is accomplished by
charging a tiny capacitor. If we consider the
two possible states of this combined system for  detection of the parallel and perpendicular
polarisation state, they do not differ appreciably in any positional coordinate of
the apparatus'
macroscopic constituents, at least not beyond the $10^{-5}$cm
given as typical dimension in collapse theories. The difference is 
due to the state of charges contributing to the current pulse which differs
as the signal traverses the electronics.  
CSL\cite{pear99_1} theories come up with collapse rates 
of $10^{-8}$s for $10^{13}$ displaced nucleons, with
the rate being proportional to the mass of the particles, and the square
of their number. The number of particles (i.e. the electrons 
contributing to the charge mentioned above) involved in the case of EPR-experiments
 is likely to be smaller, they are lighter than nucleons, and they are 
not significantly displaced. It is therefor not straightforward to assume
collapse times of less than $10^{-6}$s\cite{weih98_1}, 
even $3 \cdot 10^{-5}$s\cite{titt01_1} may be problematic.
While parameters in collapse theories can be adjusted to yield faster
rates, this introduces other measurable effects like spontaneous
temperature rises\cite{pear99_1} which have not been observed. Therefore, 
even if action-at-a-distance occurs during
measurement of entangled states, there is little reason 
to believe in superluminal speeds.

\section{Conclusions}

We have presented a model that accounts for violation of Bell type
inequalities for a classical local system. The nonlocality leading
to the violation is  outside the system and comprises the unavoidable 
connection of the two
wings of the experiment by the observer comparing the two outcomes.
The model relates to interpretations of quantum mechanics
without collapse of the wave function, or with observer mediated
collapse. 

As a further application, collapse may be introduced at various stages.
To preserve the statistics, results of both systems need to 
be transmitted to the location of the collapse, but not the
particular settings of the measurement devices. 
It is argued that the results
of current EPR experiments does not require this 
transmission to be fast.


\end{document}